\documentclass[a4paper,fleqn,usenatbib]{mnras}

\usepackage{newtxtext,newtxmath}

\usepackage[T1]{fontenc}
\usepackage{ae,aecompl}
\usepackage[version=3]{mhchem}

\usepackage{graphicx}	
\usepackage{amsmath}	
\usepackage{amssymb}	
\usepackage[usenames,dvips]{color}

\def\chandra{{\it Chandra}}
\def\xmm{{\it XMM-Newton}}
\def\sdssj01{J0100+2802}
\def\aox{$\alpha_{\text{ox}}$}

\title[XMM-Newton observation of SDSS J0100+2802]{XMM-Newton observation of the ultraluminous quasar SDSS J010013.02+280225.8 at redshift 6.326}

\author[Y. L. Ai et al.]{
Yanli Ai$^{1}$\thanks{E-mail:aiyanli@mail.sysu.edu.cn},  A.C. Fabian$^{2}$, Xiaohui Fan$^{3}$, S.A. Walker$^{2,4}$, 
G. Ghisellini$^{5}$, T. Sbarrato$^{6}$, 
\newauthor Liming Dou$^{7}$,  Feige Wang$^{8}$, Xue-Bing Wu$^{9,8}$, Longlong Feng$^{1}$
\\
$^{1}$School of Physics and Astronomy, Sun Yat-Sen University, Guangzhou 510275, China, aiyanli@mail.sysu.edu.cn\\
$^{2}$Institute of Astronomy, Madingley Road, Cambridge CB3 0HA, UK\\
$^{3}$Steward Observatory, University of Arizona, 933 North Cherry Avenue, Tucson, AZ 85721, USA\\
$^{4}$Astrophysics Science Division, X-ray Astrophysics Laboratory, Code 662, NASA Goddard Space Flight Center, Greenbelt, MD 20771, USA\\
$^{5}$INAF -- Osservatorio Astronomico di Brera, via E. Bianchi 46, I-23807 Merate, Italy\\
$^{6}$Dipartimento di Fisica ``G. Occhialini", Universit\`{a} di Milano -- Bicocca, Piazza della Scienza 3, I-20126 Milano, Italy\\
$^{7}$Center for Astrophysics, Guangzhou University, Guangzhou 510006, China\\
$^{8}$Department of Astronomy, School of Physics, Peking University, Beijing 100871, China\\
$^{9}$Kavli Institute for Astronomy and Astrophysics, Peking University, Beijing 100871, China
}

\date{Accepted XXX. Received YYY; in original form ZZZ}

\pubyear{2017}

\begin{document}
\label{firstpage}
\pagerange{\pageref{firstpage}--\pageref{lastpage}}
\maketitle

\begin{abstract}
  A brief \chandra\ observation of the ultraluminous quasar, SDSS
  J010013.02+280225.8 at redshift 6.326, showed it to be a relatively
  bright, soft X-ray source with a count rate of about 1 ct/ks. In this
  paper we present results for the quasar from a 65~ks \xmm\
  observation, which well constrains its spectral shape. The quasar is
  clearly detected with a total of $\sim$ 460 net counts in the 0.2-10 keV
  band. The spectrum is characterised by a simple power-law model with
  photon index of $\Gamma = 2.30^{+0.10}_{-0.10}$, and the intrinsic 2-10
  keV luminosity is $3.14\times10^{45}$ erg $\text{s}^{-1}$. The 1
  $\sigma$ upper limit to any intrinsic absorption column density is
  $N_{H} = 6.07\times 10^{22} {\text{cm}}^{-2}$.
  No significant iron emission lines were detected. We derive the 
  X-ray-to-optical flux ratio \aox\  of $-1.74\pm$0.01,
  consistent with the values found in other quasars of comparable ultraviolet
  luminosity. We did not detect significant flux variations either in
  the \xmm\ exposure or between \xmm\ and \chandra\ observations,
  which are separated by $\sim$ 8 months.  
  The X-ray observation enables the bolometric luminosity to be 
  calculated after modelling the spectral energy distribution: the
  accretion rate is found to be sub-Eddington.
\end{abstract}




\section{Introduction}
SDSS J010013.02+280225.8 (hereafter \sdssj01)
is an ultraluminous quasar at redshift of 6.326, which has an optical
and infrared luminosity several times greater than any other high
redshift quasars and is inferred to host a 10$^{10}$ M$_{\sun}$ black
hole \citep[][]{wu15}. The quasar is clearly detected in the
exploratory \chandra\ observation with exposure of 14.8~ks, found to
have a steep spectrum with $\Gamma$ = 3.03$^{+0.78}_{-0.70}$ derived
from the detected 14 counts \citep[][]{ai16}.  This super-massive
black hole might be growing with rapid accretion, as the bolometric
luminosity yielded from X-ray to near-infrared observations close to
the Eddington luminosity \citep[][]{wu15, ai16}.
With the peculiar properties among all quasars
discovered at $z\gtrsim 5$, which are powerful probers of cosmic
reionization \citep[][]{fan06}, \sdssj01\ sets the tightest
constraints on models for massive black hole growth and evolution at
early epochs \citep[e.g.][]{shankar09, volonteri10}.

In \chandra\ observation, the X-ray-to-optical flux ratio of \sdssj01\ is at upper envelop of the observed \aox\ values at the 
comparable ultraviolet luminosity, reported in the  Erratum to that paper \citep[][]{ai17}.
Quasars are of known to be variable and
it is quite possible that this one has been caught in a bright
state. The z=7.1 quasar, ULASJ1120+0641, is claimed to decrease in
brightness by a factor of 4 between \chandra\ and \xmm\ observation
\citep[][]{page14}, although debate exists
\citep[][]{moretti14}. There are hints of variation of \sdssj01\
during the \chandra\ exposure, which is quite puzzling if no
significant beaming effect evolved.  For high redshift quasars
extended X-ray lobes may be produced via Comptonization of cosmic
microwave background (CMB) if relativistic electrons exist
\citep[][]{fabian14}.

We proposed for a \xmm\ Director's Discretionary Time (DDT)
observation of \sdssj01 which would yield an improved spectrum with
greatly reduced errors on the spectral index, and enable a search for
any spectral features. Comparison of the flux with that from \chandra\
would provide a check on variability. Extended lobes produced from
inverse Compton scattering of CMB, which may extend over arcmin
scales, could  be detected with \xmm. In this letter we report the
spectral properties of this ultraluminous quasar from the \xmm\
observation.  Throughout this paper, we adopt the ${\Lambda}CDM$
cosmology parameters from Planck Collaboration (2014): $\Omega_{M}$ =
0.315, $\Omega_{\Lambda}$ = 0.685, and $H_{0}$ = 67.3 km s$^{-1}$.  We
define power law photon index $\Gamma$ such that N(E) $\propto$
E$^{-\Gamma}$.  For the Galactic absorption of SDSS J0100+2802, which
is included in the model fitting, we use the value of $N_{H}$ = 5.82
$\times$ $10^{20}$ cm$^{-2}$\citep[][]{kalberla05}.  All uncertainties
are given at 1$\sigma$, unless otherwise specified.

\section{XMM-Newton observation and data reduction} 
\sdssj01\ was observed with \xmm\ on 2016 June 29 for 
65 ks of Director's Discretionary Time. The European Photon Imaging
Camera (EPIC) was operated in full-frame mode, with thin filters.
The data were processed using the Science Analysis System (SAS)
version 15.0.0. The time intervals of high flaring backgrounds
contamination were identified and  excluded by inspection
of the light curves in the 10-12 keV energy range. The total cleaned
exposure times are 50 and 60 ks for the PN and MOS cameras,
respectively.  Event patterns 0-12 were included in the MOS cameras,
while for the PN camera we used patterns 0-4.  We constructed the images
in five bands, 0.2-0.5 keV, 0.5-4 keV, 4-7 keV, 7-10 keV, and then
applied source detection simultaneously using the standard
SAS task EDETECT\_CHAIN.

We extract a spectrum of \sdssj01\ from a 16$\arcsec$ radius region
around the target in each EPIC detector. The source-extraction region
corresponding to 60\%-70\% of the encircled energy fraction.  The
background was extracted from an adjacent source-free region with a
larger radius. The spectra of the target from PN/MOS cameras were
combined to form a single spectrum, with corresponding background
spectra and response matrices also combined to form a single
background spectrum and response matrix, with SAS task
epicspeccombine.  The EPIC spectra are then grouped in a way that
there are at least 25 counts in each energy bin.  We only focus on
spectrum analysis in this paper.

\section{Results} 
\begin{figure}
\includegraphics[width=0.45\textwidth]{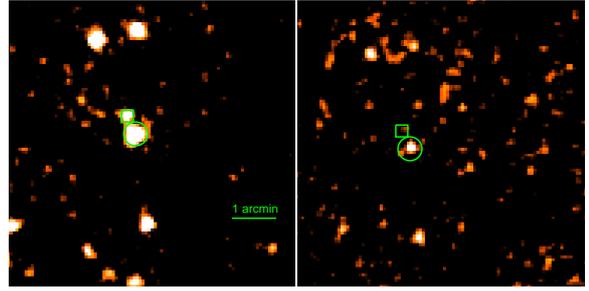} 
\caption{The 2$\arcsec$ kernel smoothed \xmm\ PN image of \sdssj01\ region of the sky in the
observed 0.3-2 keV (left panel) and 2-10 keV (right panel). The circle indicates the radius used to extract
the spectrum, and the square indicates the location of the nearby X-ray
source SDSS J010013.95+280250.6.
\label{fig1}}
\end{figure}

As shown in Figure\,\ref{fig1}, \sdssj01\ is clearly detected in the
\xmm\ EPIC images.  The most accurate source position, from the VLBA
1.5 GHz image, lies within the astrometric uncertainties of both the
optical Sloan Digital Sky Survey and the \chandra\ X-ray observation
\citep[][]{wang17}.  The XMM X-ray position of the quasar given by SAS
task EDETECT\_CHAIN is $\sim$ 1.7 arcsec away from the radio poistion,
with a 1$\sigma$ position uncertainty of 0.6 arcsec.

The detected net counts of \sdssj01\ in 0.2-10 keV is 460.  
\sdssj01\ is relatively soft with weak detection in hard X-ray band, 2-10 keV (Figure\,\ref{fig1}). 
It is detected individually in 0.2-0.5 keV and
0.5-4 keV with false probability less than $10^{-10}$; While, in 4-7
keV the detection significance of the quasar is close to 3 sigma with
a false probability of 0.015. It is not detected in the 7-10 keV band. 

The nearby X-ray source, SDSS J010013.95+280250.6, which is detected
in \chandra\ observation 28$\arcsec$ to the northeast of \sdssj01, is
also detected in the \xmm\ EPIC image (Figure\,\ref{fig1}). This
object is relatively faint in X-ray emission with detected net counts of
80 in 0.2-10 keV within a 15$\arcsec$ radius aperture in the EPIC images.
It is not detected in the hard X-ray band (2-10 keV in the observed
frame) with an upper limit of 10$^{-4}$ cnt s$^{-1}$ estimated from
the sensitivity maps using the SAS task {\it esensmap} for a
logarithmic likelihood of 12.  According to the Point Spread Function,
the counts from this faint object, which fall in the source extraction
region of our target quasar, are $\sim$ 8 counts. Our target quasar
therefore has little contamination from its emission.

The image of J0100+2802 appears slightly lop-sided to the South East, as shown in Figure 2.
The excess flux in 0.5-2 keV is about $2\times 10^{-16}$ erg s$^{-1}$ cm$^{-2}$. 
A deep Chandra image is required to distinguish several unresolved faint point sources from 
possible diffuse inverse Compton emission. If it is the latter due to a jet from the quasar then 
it may be detectable in the radio band below the mJy level.

\begin{figure}
\includegraphics[width=0.45\textwidth]{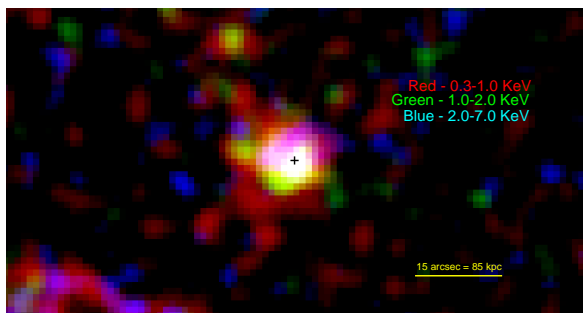} 
\caption{A RGB color image of \sdssj01\ using different bands from the EPIC PN image. 
Red shows soft X-ray emission (0.3-1.0 keV), green shows intermediate emission (1.0-2.0keV) and 
blue shows the hard X-ray emission.
\label{fig_rgb}}
\end{figure}

\begin{figure}
\includegraphics[width=0.35\textwidth,
angle=-90]{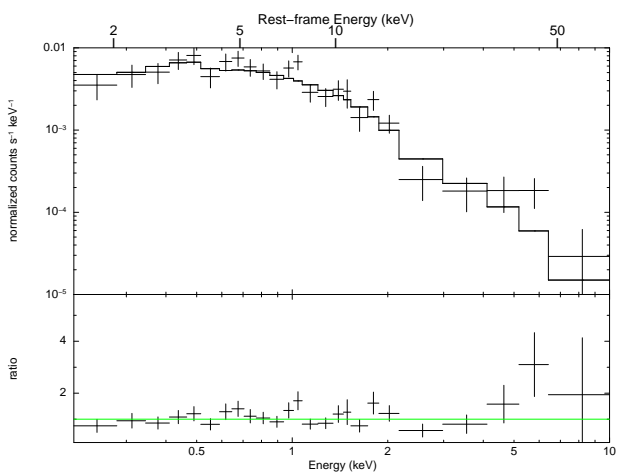} 
\caption{Upper panel: \xmm\ spectrum of \sdssj01\ and power-law model
with fixed Galactic absorption. Lower panel: ratio of the data to model. The relatively
larger data to model ratio at energies greater than 4 keV maybe due to
dominance of X-ray background emission above 4 keV for this quasar.
\label{fig2}}
\end{figure}

We fitted the spectrum of \sdssj01\ using XSPEC \citep[v12.9;][]{arnaud96}
using a simple power-law model modified by Galactic absorption. 
The fitted photon index is $\Gamma = 2.30^{+0.10}_{-0.10}$. The
 fit is acceptable, with a $\chi^{2} = 24.7$ for 23 degrees of freedom
 (Figure\,\ref{fig2}).  We also fold the model with intrinsic
 absorption (at z=6.326). There is no significant improvement with
 $\Delta\chi^{2}$ of 1.7, and 1 $\sigma$ upper limit of the intrinsic
 absorption column density is
 $N_{\rm H} = 6.07\times 10^{22} {\text{cm}}^{-2}$.  There are
 residuals at energy between 5--10 keV, as shown in
 Figure\,\ref{fig2}, which are possibly due to contamination from
 background as the source detection significance in this energy range
 is below 3 sigma. The rest-frame 2-10 keV luminosity implied by the
 fit is 3.14$^{+0.53}_{-0.48}$$\times10^{45}$ erg $\text{s}^{-1}$.

No Fe K emission line feature appears to be present in the residuals,
and the 1$\sigma$ upper limit for the iron K$\alpha$ equivalent width
is 0.02 keV (rest frame).
There are relatively
larger data to model ratio at energies greater than 5 keV (rest frame $\sim$ 36keV), which are
possibly due to the contamination from statistical poisson fluctuation
of the background emission. As shown above, the detected
significance of \sdssj01\ at 4-7 keV is only at the level of
3$\sigma$.  Further deep exposures can help to justify whether
the spectral shape of this quasar deviates from a simple power-law at
high energies.

The light curve for \sdssj01\ is extracted and no significant
variation is detected during the \xmm\ exposure.  We then compare the
X-ray spectrum and flux between the \chandra\ and \xmm\ observations
with time interval of about 8 months. First, the value of the inferred
photon index from \xmm\ observation is within the errors of the one
from \chandra\ observation, which is $\Gamma =
3.03^{+0.78}_{-0.70}$. That is, no statistical spectral shape
variation was detected between the two observations for
\sdssj01. Second, there is none detection of
flux variation between the two observations, with the rest-frame 2-10 keV luminosity implied by the
fit in \xmm\  observation consistent within errors with the one, 9.0$^{+9.1}_{-4.5}$$\times10^{45}$ erg $\text{s}^{-1}$, from \chandra\ observation.
Finally, the
residual at $\sim$ 1.2 keV, hinted at in the \chandra\ spectrum of
\sdssj01, was not detected. The non-detection in \xmm\ observation
indicates that the feature in \chandra\ spectrum was probably due to
instrumental lines \citep[][]{bartalucci14}, 
although Poisson fluctuation can not be excluded.

\section{Discussion}

\sdssj01\ is significantly detected in the \xmm\ observation with
total net counts of 460 in the 0.2-10 keV band.  A simple power-law model
provides acceptable fits to the spectrum with inferred photon index of
$\Gamma = 2.30^{+0.10}_{-0.10}$. The value of $\Gamma$ is consistent with the one found by \citet[][]{nanni17}.
The 1 $\sigma$ upper limit on any
intrinsic absorption column density is
$N_{H} = 6.07\times 10^{22} {\text{cm}}^{-2}$.  No significant iron
emission lines were detected. With the well constrained X-ray spectral
shape and luminosity, we now discuss the emission from accretion disk with
broad-band energy spectral analysis for \sdssj01, and compare the
spectral energy distributions (SED) of this quasar with other high-redshift and
low-redshift quasars.

\subsection{Black hole mass and disc luminosity of J0100+2802}
The black hole mass estimated by \citet[][]{wu15} is $M_{\rm BH}=1.2\times 10^{10} M_\odot$.  
This is based on the virial method,
and it is therefore affected by an uncertainty of a factor 3
\citep[acknowledged by][]{wu15}.  The bolometric luminosity, assumed
isotropic, given by \citet[][]{wu15}.  is $L_{\rm bol}=1.6\times 10^{48}$ erg
s$^{-1}$, and includes the infrared and the X--ray emission \citep[following][]{shen11}.  
The corresponding optical--UV emission is nearly 1/2 of
that \citep[][]{calderone13}.  The other half is reprocessed emission
in the infrared by the absorbing torus surrounding the disc, plus the X--ray
emission produced by the corona sandwiching the disc.  The latter
could indeed be energised by the gravitational energy of the accreting
matter.

Both the black hole mass and the accretion luminosity are huge, and
motivate us to explore alternative methods to reliably measure them.
A proper accretion luminosity estimate should exclude the infrared
reprocessed emission and take into account possible anisotropies.  We
therefore use a standard \citet[][]{shakura73} accretion disc
model to fit the observed optical--UV flux,
while the corona X--ray emission is treated phenomenologically
by adding a power law starting below the peak of the disk emission,
ending with an exponential cut. 
This component requires three parameters: normalization, slope and
cut frequency.

We are aware of the limitations connected with the use of the Shakura-Sunyaev 
disk model, mainly due to i) the spin is assumed to be zero; ii) all relativistic effects 
are neglected and iii) the disk is assumed to be geometrically thin and optically thick.
The first assumption would lead to a {\it lower limit} on the black hole mass
and to an {\it upper limit} on the accretion rate, as discussed below.
The second assumption introduces an uncertainty on the angular pattern of the produced
radiation, but not on the overall shape of the spectrum \citep[see, e.g. ][]{campitiello17}.
The latter assumption is questionable in the case of near (or above) Eddington accretion,
since because the disk could become geometrically thicker close to the black hole.

Assuming a null spin implies an innermost radius of the circular orbit
($R_{\rm ISCO}=6R_{\rm g}$, $R_{\rm g}$ is the gravitational radius)
and a corresponding accretion efficiency  (defined by $L=\eta \dot M c^2$)
equal to 0.057 or to 0.08 according if relativistic effects are included or not.
By increasing the spin, $R_{\rm ISCO}$ decreases, to become $R_{\rm g}$ when
the dimensionless spin $a\sim 1$.
Correspondingly, $\eta$ increases, reaching a theoretical maximum of 0.42,
that is however reduced to $\eta=0.32$ \citep[][]{thorne74} when properly
including the effects of accretion (and of photons produced by the disk 
falling into the black hole).
The black hole spin has a negligible effect on the outer regions of the disk
emitting in the infrared--optical band,
but changes the emitting properties of the inner radii.
In other words, for a given accretion rate and black mass, the disk around
a rotating hole will produce the same amount of IR radiation, but more UV 
than a Shakura--Sunjaev disk.

The SED of J0100+2802 shows indeed a peak, allowing to  find the total luminosity,
associated to the accretion rate for the assumed efficiency $\eta$,
and the black hole mass, since the peak frequency is associated to the
temperature of the innermost orbits contributing to the observed spectrum.
Applying the Shakura--Sunjaev model (i.e. zero spin) we then find $M$ and $\dot M$.
If we assume a non--zero and positive spin, the total luminosity
can be produced with a reduced accretion rate ($\eta$ is larger), but
this implies that we underestimate the flux in the optical--IR bands
(flux produced at larger radii).
Therefore we have to increase the black hole mass (and therefore the surface of the disk)
to make the disk ``colder"  in order to fit the entire spectrum.

We consider the
anisotropic emission of the disc, that follows a pattern
$\propto\cos\theta$, 
and assume that object is observed under a
viewing angle 30$^\circ$ from the disc's normal (i.e.\ the average
angle between 0$^\circ$ and an assumed aperture angle of the torus of
45$^\circ$).  
Along with the disc emission, we assume a blackbody
emission at a temperature $T_{\rm torus}$ to model the torus emission.
For the corona X--ray component, as explained above, we assume
a power law of photon index $\Gamma=2.5$ ending with an exponential cut ($h\nu_{cut}=300$ keV), 
emitting a fraction $L_x/L_{\rm disc}\sim 1/3$ of the optical--UV luminosity.  
The infrared and X--ray fluxes are assumed to
be emitted isotropically.

\begin{figure}
\includegraphics[width=0.5\textwidth]{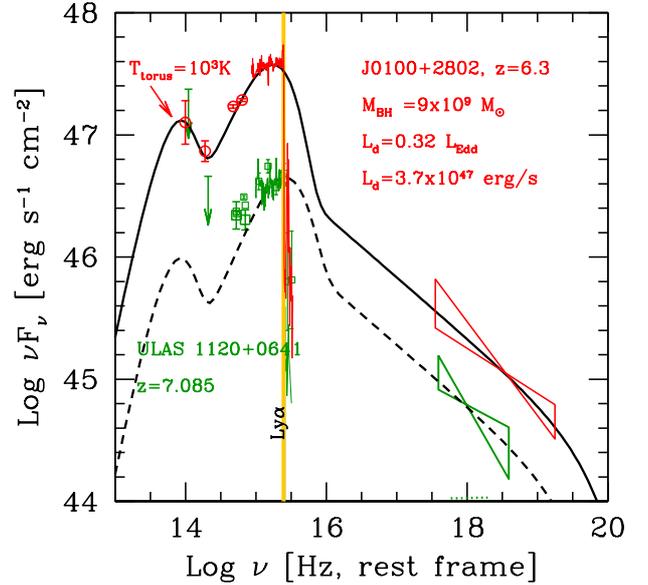} \vskip -0.5
cm
\caption{ The spectral energy distributions of J0100+2802 (symbols in red)
and our fitting model (solid black line) compared to the SED (symbols in green) 
and model (dashed black line) of ULAS J1120+0641. The vertical orange line labels the Ly$\alpha$ line.
The inferred black hole mass and accretion luminosity for J0100+2802 are
indicated.  Infrared data are from WISE, optical spectra from the works by
\citet[][]{mortlock11} and \citet[][]{wu15}, respectively.  X--rays of
J0100+2802 are from this work.  } \vskip 0.5 cm
\label{1120}
\end{figure} 

\begin{figure}
\includegraphics[width=0.45\textwidth]{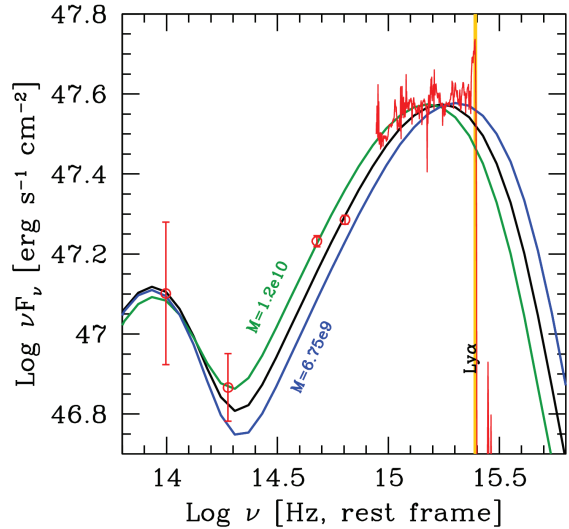} \vskip -0.5cm
\caption{SED modelling of \sdssj01, corresponding to the same total luminosity, but with three different 
masses (solid black line is the one with mass of 9$\times 10^9 M_\odot$).} \vskip 0.5 cm
\label{0100zoom}
\end{figure}

Figure \ref{1120} shows the infrared to X--ray spectral energy distributions (SED) of J0100+2802, together
with the fitting model.  
The disc optical--UV luminosity is $L_{\rm
disc}\sim 3.7\times 10^{47}$ erg s$^{-1}$, that corresponds to 32\% of
the Eddington luminosity, for a black hole mass of $M_{\rm BH}=
9\times 10^9 M_\odot$, slightly smaller than the estimate of \citet[][]{wu15}, but still consistent.  The total X--ray luminosity
(i.e. from the peak frequency of the disc emission to $\sim$1 MeV) is
$\sim$ 1/3 of $L_{\rm disc}$.  This gives $L_X+L_{\rm disc}\sim 5
\times10^{47}$ erg s$^{-1}$, equivalent to $0.43 L_{\rm Edd}$. 
As explained above, the assumption of zero spin, implicit 
in the use of the Sakura--Sunjaev model, implies that the
derived value of the black mass is a lower limit.
This strongly suggests that the disk luminosity, including the
rather large X--ray component, is  sub--Eddington.
The uncertainty of the derived black hole mass is $\sim$ 0.4 dex, as shown in Figure\,\ref{0100zoom}. 
In the figure we show the SED modelling of \sdssj01, corresponding to the same luminosity, but with different masses.

\subsection{Comparison with ULAS J1120+0641}
Figure\,\ref{1120} includes the SED of ULAS 1120+0641, the quasar with
the largest measured redshift \citep[$z=7.085$;][]{mortlock11}.  In
the far--infrared band we have only upper limits to the flux, that are not
very constraining.  Note also some discrepancy between the photometric and the spectral data 
at the same frequencies. 
For the fit, we have given priority to the spectroscopic data.

This source is less luminous than J0100+2802, and
its mass is smaller, according to the estimate obtained by fitting the
SED.  With the same accretion disc model as before, in fact, we obtain
$M_{\rm BH} =1.3\times 10^9 M_\odot$; $L_{\rm disc}= 3.9\times
10^{46}$ erg s$^{-1} =0.23 L_{\rm Edd}$; $L_X =0.8L_{\rm disc}\sim
3\times 10^{46}$ erg s$^{-1}$ and $L_X +L_{\rm disc} =0.41 L_{\rm
Edd}$.  As previously explained for J0100+2802, the value  of the black hole mass should be taken as a lower limit.
We conclude that both sources, despite the difference of black
hole mass, share similar Eddington ratios and similar partition
between optical--UV and X--ray luminosities.  We can compare our
results on ULAS J1120+0641 with the ones of \citet[][]{mortlock11} who
found $M_{\rm BH}\sim 2\times 10^9 M_\odot$ (through the virial
method) and a disc luminosity of $2.5 \times 10^{47}$ erg s$^{-1}$
(applying a a fiducial bolometric correction taken from \citet[][]{willott10}.  
Differently from us, the results of \citet[][]{mortlock11} indicate a slightly super--Eddington luminosity.

\subsection{Comparison with other powerful quasars}

\begin{figure}
 \includegraphics[width=0.45\textwidth]{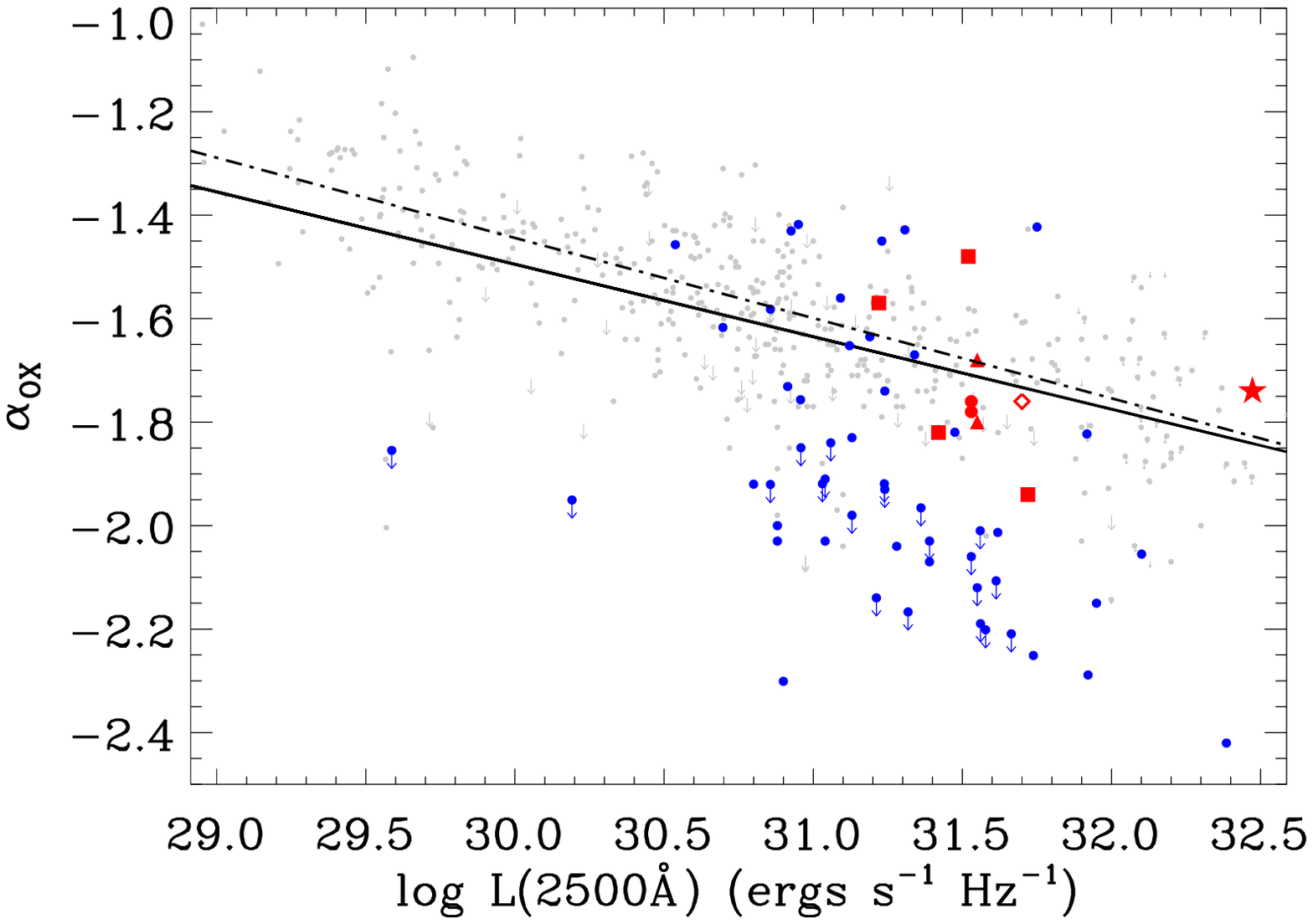} \vskip 1 cm
\caption{Location of \sdssj01\ (red star) in the X-ray-to-optical
power-law slope parameter \aox\ vs. 2500 \AA\ monochromatic
luminosity. The grey dots are the quasars from the samples of \citet[][]{just07, steffen06} and \citet[][]{gibson08}. The blue
dots are the weak line quasars and PHL 1811 analogs from
\citet[][]{luo15}. The solid line represents the \aox-$L_{2500 \text \AA}$ relation
from \citet[][]{just07} and the dotted-dashed line from \citet[][]{nanni17}. 
The red symbols represent the high-redshift quasars
with z$>$6.0 from literatures (squares from \citet[][]{shemmer06} , filled circles of ULAS J1120+0641 \citep[][]{moretti14, page14}, 
triangles of SDSSJ1030 \citep[][]{brandt02, farrah04}, and diamonds of SDSSJ1148+5152 \citep[][]{gallerani17}).
\label{fig4}}
\end{figure}

It is well established that the X-ray-to-optical power-law slope parameter \aox\ of quasars significantly 
correlate with the ultraviolet 2500\AA\ monochromatic luminosity \citep[$L_{2500 \text\AA}$,][]{steffen06, just07}.  
For \sdssj01\, with rest frame 2500\AA\ flux
density, $f_{2500\text{ \AA}}$, estimated from Wu et al. (2015) and
rest-frame 2 keV flux density, $f_{2\text{ keV}}$, estimated from the
power law model, we have the parameter \aox\ of -1.74$\pm$0.01. 
In Figure\,\ref{fig4}, we show the location of \sdssj01\  and the other highest-redshift quasars with z$>$6, of which we take the
\aox\ and $L_{2500 \text \AA}$ from literatures
\citep[][]{brandt02, farrah04, shemmer06,moretti14,page14,gallerani17}, in the \aox--$L_{2500 \text\AA}$ relation.
It is clear that the SED of the ultra-luminous \sdssj01\ is not abnormal among the highest redshift quasars 
and all the quasars at z$>$6 follow the  \aox--$L_{2500 \text\AA}$ relation as the low- and median-redshift quasars.
As discussed in \citet[][]{nanni17}, which presents a systematic analysis of X-ray archival data of quasars at $z>5.5$, these
results support the non-evolutionary scenario of the SEDs of luminous quasars. 
For \sdssj01\ the inferred value of \aox\ in \citet[][]{nanni17} is  -1.88$^{+0.01}_{-0.02}$, which is in agreement
with ours considering the scatter of the \aox--$L_{2500 \text\AA}$ relation. 

With rest-frame equivalent width of the Ly$\alpha+\ce{Nv} \sim 10 \text\AA$ \citep[][]{wu15}, \sdssj01\ is one 
of the Weak-line quasars (WLQs), which are a subclass of radio-quiet quasars that have almost extremely weak 
or undetectable emission lines \citep[e.g.][and references therein]{fan99,meusinger14}. Significant 
fractions ($\sim50\%$) of the WLQs are distinctly X-ray weak compared to typical quasars \citep[][]{shemmer09,wu12,luo15}. 
While, as shown in Figure\,\ref{fig4}, \sdssj01\ is not X-ray weak, compared to the SEDs of the other WLQs.  \sdssj01, presented 
as an X-ray normal weak-line quasar, provides constraints about the proposed hypotheses to the
interpretation of weak-line quasars, such as a soft ionizing spectral energy distribution due to intrinsic 
X-ray weakness or due to small-scale absorption \citep[e.g.,][]{leighly07, wu12,luo15}.

We did not detect variation of the X-ray emission for \sdssj01\ in the
\xmm\ exposure, and no signifiant variation was detected in the X-ray
flux observed from \xmm\ and \chandra\ observations. For this high-redshift radio-quiet luminous quasar, 
the non-detection of variation is not un-expected. Also, the results
normally rule out the possibility from jet beaming effect in the
observed X-ray brightness of \sdssj01, in which case there should be  
detected variations.

\section{Summary}
With the well-detected X-ray emission from \xmm\ observation,
\sdssj01\ presents as an peculiar high-redshift quasar in X-ray with
relatively soft X-ray spectral shape. 
With the X-ray observation, the bolometric luminosity is
calculated from spectral energy distributions modelling and the
accretion rate is estimated to be sub-Eddington.  
The location in the \aox-$L_{2500 \text \AA}$ relation indicates it is an 
X-ray normal quasar either in term of high-redshift quasars or weak-line quasars.
The results from
\xmm\ observation of \sdssj01\ are meaningful for the study of quasar
X-ray properties, broad-band energy distribution, and super-massive
black hole formation and evolution at cosmic dawn.

\section*{Acknowledgements}
Y.-L.A. and L.L.F. acknowledge the support from the NSFC grants
11273060, 91230115 and 11333008, and State Key Development Program for
Basic Research of China (No. 2013CB834900 and 2015CB857000).
A.C.F. and S.A. W. acknowledge support from ERC Advanced Grant 340442.
S.A.W. was also supported by an appointment to the NASA Postdoctoral
Program at the Goddard Space Flight Center, administered by USRA
through a contract with NASA.  X.F. acknowledges support from NSF
grant AST 15-15115 and from the Institute of Astronomy, University of
Cambridge through a Raymond and Beverly Sackler Distinguished Visitor
program.  F.W. and X.-B.W. acknowledge the support from NSFC grants 11373008 and 11533001.
We thank Tinggui Wang, and Junxian Wang for helpful
discussions.  This work is based in part on observations obtained with
\xmm, an ESA science mission with instruments and contributions
directly funded by ESA Member States and NASA. We thank Dr Norbert
Schartel for the allocation of XMM observing time.











\bsp	
\label{lastpage}
\end{document}